\documentstyle[12pt]{article} 
\begin{document} 
\centerline{\LARGE { Solitons and wavelets: Scale analysis and bases}}
\vskip 1cm 
\centerline{A. Ludu, R. F. O'Connell and J. P. Draayer} 
\centerline{Department of Physics and Astronomy,} 
\centerline{Louisiana State University, Baton Rouge, LA 70803-4001} 

\begin{abstract}
We  use a one-scale similarity analysis which gives specific relations between
the velocity, amplitude and width of localized solutions of nonlinear differential
equations, whose exact solutions are generally difficult to obtain.  
We also introduce  kink-antikink compact solutions  for the nonlinear-nonlinear dispersion
K(2,2) equation, and we construct a basis of scaling functions similar with those used in the
multiresolution analysis.  These approaches are  useful in describing
nonlinear structures and patterns, as well as in the derivation of the time evolution of
initial data for nonlinear equations with finite wavelength soliton solutions.
\end{abstract}
\vskip 0.8cm
PACS numbers: 42.30.Sy, 02.30.Px, 43.35.+d, 47.20.Ky, 05.45.-a,

\vskip 3cm
\section{Introduction}

The nonlinear partial differential equations (NPDE) of physical interest can describe
a variety of patterns, particle-like traveling solutions, solitons and breather modes 
in nuclear
and particle physics, nonlinear molecular and solid state physics phenomena, and
features found in nonlinear optics \cite{general1}. Their solutions are usually localized and
demonstrate stability in time and in collisions with each other. In the asymptotic domain
these solutions consist of isolated traveling pulses  that are free of interactions and
have a shape  related to the velocity, thus making nonlinear patterns distinct from linear
results.   In the scattering domain, the nonlinear solutions obey some nonlinear superposition
principle.  Nonlinear dynamics require  NPDE which display very
strong interaction between the initial conditions and the dynamics and  involve
multiple scales \cite{chaos,multiple}, being able to  produce self-similar or fractal-like
patterns. Recent
examples show that the traditional nonlinear tools (inverse scattering, group symmetry,
functional transforms) are not always applicable \cite{compacton1}. From the experimental point
of view one knows that such patterns generally have finite space-time extension and a
multi-scale structure. Since the traditional solitons or the soliton-like solutions  have 
infinite extent, one needs rather appropriate compact supported basis functions to
investigate such structures. 

Multi-resolution analysis (MRA), \cite{wavelet1}, could be a useful method for the construction
of such nonlinear bases, since  the  linear harmonic analysis is inadequate
for describing nonlinear systems. The MRA uses wavelets, which are functions that
have a space-dependent scale which renders them an invaluable tool for analyzing multi-scale
phenomena. Wavelets have been used in signal processing, problems
involving singular potentials in quantum mechanics, in discussions concerning q-algebras,
and even in nuclear structure studies \cite{wavelet2}. It follows that the use of MRA
in the study of NPDE is natural because wavelets  can analyse nonlinear features like
strong variations or singularities. The NPDE describe 
a variety of patterns \cite{chaos}, features in quantum optics \cite{optics}, 
molecular and solid state physics phenomena \cite{molecular}  and 
solitons in  nuclear and particle physics \cite{nuclear,particle}.

In this paper, wavelet-inspired approaches for localized solutions of NPDE are explored.
We propose two different formalisms for the scale analysis, and
the classification of soliton solutions of NPDE. A first method provides
relations between the characteristics of such solutions (amplitude, width and velocity) 
without the need of solving the corresponding NPDE. The method uses the
multi-resolution analysis
\cite{wavelet1} instead of the  traditional tools like the Fourier integrals or linear harmonic
analysis which are inadequate for describing such systems.  Moreover, the
introduction of wavelet analysis in the study of NPDE is somehow natural because it can
accommodate behaviour diplaying  strong variations, even singularities, to a smooth behavior.
This scale approach has the advantage that it does not need the explicit form of the exact
solutions. Hence, it is useful especially in  situations when such solutions are unknown. In
section 2 we provide  many  examples, predictions and applications of this one-scale
approximation method (OSA) for  a large class of NPDE with respect to their localized solutions.
The NPDE is mapped into an algebraic equation relating the amplitude, width and velocity of such
signals. The results are succinctly presented in tabular form. In  section 2 we show an
example of the construction of a nonlinear basis for  NPDE i.e.  the construction of a
kink-antikink basis for the K(2,2) equation, involving nonlinear dispersion.  There are many
physical reasons favoring wavelets in the construction of such nonlinear bases. The
self-similar character of the fission process of fluid drops is an example where the same type of
singularity  occurs in any scale \cite{chaos,prl,bona}.  In section 3, we use also this basis (or
frame) for the investigation of the time evolution of a given initial data profile for the
nonlinear-nonlinear dispersion K(2,2) equation. The proofs are introduced in two Appendices.

\section{Scale analysis of NPDE}

In this section we introduce a one-scale analysis (OSA) for the NPDE, in terms of their
localized traveling solutions. The NPDE which describe physical phenomena are usually 
not susceptible to analytic solutions. Moreover, there are examples \cite{compacton1,fred}
when the mathematical tools like the inverse scattering theory or the transformation group 
method are not  applicable. When  phenomena of interest have many space-time scales,
or the scale of the process varies in time, the numerical methods may fail, like in
the case of  propagating  discontinuities or shock waves. A simple option is the expansion of
solutions in a basis of appropriate chosen basis. The  
Fourier series, which has the advantage of orthogonality,  cannot discriminate the  local
behavior of phenomena.  Moreover, the analysis of the Fourier
coefficients of a wave $u(x,t)$ is not sufficient for drawing conclusions about the
scale of the localized structures.

The so called OSA analysis, described and applied here for  localized 
traveling solutions belonging to any type of NPDE, provides  algebraic connections between the
width  $L$,  amplitude $A$, and the velocity $V$ of the solution, without actually solving the
equation. The procedure consists in the  substitution of all the  terms in the NPDE, according
to the rules: 
$$
\hbox{i.} \ \ u_t \rightarrow -Vu_x \ \ \ \hbox{substitution for traveling solutions},  
$$
\begin{equation} 
\hbox{ii.} \ \ \ u \rightarrow \pm A, \ \  \ \ \
u_x \rightarrow \pm A/L,  \ \ \ \ \ 
u_{xx} \rightarrow \pm A/L^2 \dots ,
\end{equation}
and so forth for higher order of derivatives. This substitution in eq.(1 ii)
is possible only for localized (finite extended support) solutions  having at least  one local
maximum (like solitons or Gauss functions), if they exist. The advantage
of the substitution follows from the fact that it enables one to say something about the
hypothetical localized traveling solutions  even if (or especially when) they are not known or
discovered. However, this averaging method may fail in the case of strictly monotonic solutions,
or solutions with singularities.  This will be made more explicit in the next paragraph and
Appendix 1.

Since we are interested in traveling solutions, the first substitution reduces the number of
variables from 2 to 1 so that we are now dealing with an ordinary differential equation
instead of a PDE. Then, the second substitution transforms the ordinary differential equation
into an  equation in the parameters describing the amplitude, width and velocity. Consequently,
the NPDE is mapped into an algebraic equation in
$A,L$ and $V$. 

The proof of the method follows from the expansion of the soliton-like 
solution $u(x,t)=u(s)$ with $s=x-Vt$, in a Gaussian family of wavelets  $\Psi (s)=Ne^{Q(s)}$,
where $Q(s)$ is a polynomial and $N$ the normalization constant \cite{wavelet1,wavelet3}.
If we choose $Q= -i s-{{s^2}\over 2}$ we obtain a very particular wavelet with the support
mainly confined in the $(-1,1)$ interval, namely $\Psi (s)= \hbox{exp}[-i
s-{{s^2}\over 2}]/\pi ^{1/4}$. We have the discrete   wavelet expansion of
$u$ 
\begin{equation}
u(s)=\sum_{j}\sum_{k}C_{j,k}\Psi  (2^j s -k)=\sum_{j,k}C_{j,k}\Psi _{j,k}(s),
\end{equation}
in terms of integer translations ($k$) of $\Psi $, which provide the analysis of localization,
and in terms if dyadic dilations ($2^j$) of $\Psi $, which provide the description of different
scales. The idea of the proof is to choose a point where the expansion in eq.(2) can be well
approximated by one single scale such that $u(s)$ can be approximated with a sum of phases in
$s$. In order to reduce the number of scales, the range of the summation
should be chosen with an eye to the underlying physics. The exact proof of this approach is
provided in Appendix 1.

In the following  we apply this procedure to three classes of NPDE, namely:
dispersive, diffusive and dispersive-diffusive nonlinear equations, and we present
the results in tabular form. 
Before describing different examples of NPDE, we point out that for each equation we try
to present one example of an exact solution (second column  Tables 1,2) in comparison with the
results of the much simplest OSA approach (third column of first two
tables and second column of Table 3). We stress that we do not choose only that solution which
fits the similarity analysis. On the contrary, our approach fits a large number of solutions, as
can be checked directly.

Our first example of application of the OSA approach is given by the convective-dispersive
equations, for which the most celebrated example is provided by the KdV equation
\begin{equation}
u_t +uu_x + u_{xxx} =0.
\end{equation}
In general, the stable solutions of the  nonlinear-dispersive
equations are dependent of the initial conditions, through their conservation laws. 
Consequently, they can generate a large class of patterns, shaped by the balance between
nonlinear interaction and dispersion, among which the most interesting examples are solitons,
breathers and  kinks. 

In Table 1 we present examples of  pure dispersive NPDE, identified in the
first column by the form of the equation, and in the second column by a 
corresponding traveling localized solution, if the analytical form is available. Such 
exact solutions provide special relations between $L,A$ and $V$, which are given in the third
column of Table 1. In the last column we introduce the results of OSA, namely the  relations
between these three parameters, provided  by  eqs.(1 ii).   The usefulness of the approach may be
checked, by a quick comparison between the second and the third columns. While the results in
the second column are possible only when one knows the analytical solutions, the results
presented in the last column, obtained by the  OSA, result directly from the
NPDE form, without actually solving it.

The case of the KdV equation, eq(3), is described in the first row of the Table 1. 
The OSA method gives a general expression for $L=L(A,V)$, shown in the first line, first row of
the last column. In  the second line of the last column we show that a specification of the
relation between $V$ and $A$, together with the general result given in the first line, results
in a relation between $L$ and $A$.  For $L$ to be  related
to $A$ only,  it results from the third column that the velocity $V$ must be proportional 
to $A$. In this case we obtain  the well-known relation (first row, column two) among
the parameters in the soliton solution $V=2A$. Moreover, OSA method allows $V$ to depend on a
higher power of $A$ ($V \sim A^{p}$, $p \geq 1$). If such a solution could exist, a lower bound
for $A$ will occur.  Such solitons would have only amplitudes higher than this limit, while
solitons with a smaller amplitude than this limit  move with velocity proportional to $A$. 

Similar results are obtained for the MKdV equation (second row), except that here $A$ needs to
be proportional to the square root of $V$ in order to have $L$ a function of $A$ only. This
prediction is again identical with that in the  exact solution (second column).  Moreover, the
same relations remain valid even for  the   solutions  of the "compacton" type 
\cite{electronic}
$$
u(x,t)={{\sqrt{32}k \cos [k(x-4k^2 t)]^2 }\over {3(1-{2\over 3} \cos [k(x-4k^2 t)]^2)}},
$$
where $L=  \pi /6 k$, that is $L\sim 1/A$, like in the Table 1.

Next example (third row) is provided by a generalised KdV equation, in which the dispersion
term is quadratic
\begin{equation}
u _t + (u ^2)_x +(u ^2)_{xxx}=0.
\end{equation}
Eq.(4), known as K(2,2) equation because of the two quadratic terms, admits compact supported
traveling solutions, named compactons \cite{compacton1,fred,rosenau1,rosenau2,rosenau3}.  In
general, the compactons are obtained in the form of a power of  some trigonometric  function
defined only on its half-period, and zero otherwise, in such a way that the solution is enough
smooth for the NPDE in discussion. In the above example the square of the solution has to be
continuous up to its third derivative with respect to $x$.

Different from solitons, the compacton width is
independent of the amplitude and this fact provides the special connection with the wavelet
bases. The compactons are characterize by a unique scale, and it is this feature
that makes it possible to introduce a nonlinear basis starting from a unique generic
function.  For eq.(4) the compacton solution is given by
\begin{eqnarray}
\eta _c (x-Vt)={{4V}\over {3}} cos ^2 \biggl [  {{x-Vt}\over {4}} \biggr ], 
\end{eqnarray}
if $|x-Vt|<2\pi $ and zero otherwise. Here we notice that the velocity is proportional to the
amplitude and the width of the wave is independent of the
amplitude, $L=4$.  As a field of application we mention that the quadratic
dispersion term is characteristic for the dynamics of a chain with nonlinear coupling.  

The general compacton solution for eq.(4) is  actually a ''dilated" version of eq.(5).
That is, a combination of the first rising half of the squared cos in eq.(5), followed
by a flat domain of arbitrary length $\lambda $, and finally followed by the second,
descending part of eq.(5). Actually, this combination is just a kink compacton joined
smoothly with an antikink one
\begin{equation}
{\eta}_{kak} (x-Vt; \lambda)=
\left\{ \begin{array}{ll}
0 ...\\
{{4V}\over 3} \  cos^{2}\biggl [ {{x-Vt}\over 4} \biggr ],
\ \   -2\pi \leq x-Vt \leq 0
\\
{{4V}\over 3} ,  
\ \   0 \leq x-Vt \leq \lambda  \\ 
{{4V}\over 3} \  cos^{2}\biggl [ {{x-Vt-\lambda }\over 4} \biggr ],
\    \lambda \leq x -Vt \leq \lambda +2\pi 
\\ 
0 ... \\
\end{array}
\right.
\end{equation}
In Fig. 1 we present compactons (upper line),   kink-antikink
pairs (KAK) described by eq.(6),  both with the same amplitude and velocity (middle line).
Although the second derivative of this generalized compacton is discontinuous at its edges, the
KAK, eq.(6),  is still a solution of eq.(4) because the third derivative acts on $u^2$, which is
a function of class $C_3$. Finally, we can construct solutions by placing a compacton on the top
of a KAK, as in the bottom line of Fig. 1. Such a solution exists only for a short interval
of time ($\lambda /V$), since the two structures have different velocities. The analytic
expression of the solution is given by
\begin{equation}
\eta (x,t)=\eta _{kak} (x-Vt;\lambda )+\biggl ( \eta _c (x-V't-2\pi) + {{4V}\over 3 }
\biggr ) \chi ({{x-V't-2\pi} \over {2\pi}} ),
\end{equation}
for $0< t < (\lambda -4\pi)/(V'-V)$ and zero in the rest. Here $\chi(x)$ is the support function,
equal with 1 for
$|x|\leq 1$ and 0 in the rest, and $V'=3\hbox{max}\{ \eta _c\}/4 +2V $.

For the K(2,2) compacton eq.(5) fulfills some relations between the parameters: 
$A=4V/3$ and $L=4$ \cite{rosenau1}. The relations provided by OSA in the last
column of the third row, predict such relations, and hence also proove the existence of the
compacton. That is, for a linear relation between the  amplitude and the speed, the half-width
is constant and does not depend on $A$.  Indeed, if we choose $V=\pm 3A/2$ or $V=\pm 5A/2$, we
obtain $L=4$, like for the compacton. The constant value of $L\equiv L_0=const.$ is a
typical feature of  the K(2,2) compactons. Moreover, it was found numerically that for any
compact supported  initial data, wider than $L_0$, the solution decomposes in time into a series
of
$L_0$ compactons, Fig. 2. For narrower initial data the numeric solution blows up. There
is no exact or analytic explanation of this effect, so far.  The scale relations can give a
hint in this situation, too, by using the graph of the relation
$L=L(V,A)$ provided by this qualitative method. In Fig. 3, $L$ from the third column of the
third row is plotted versus $V$, for several values of $A$ (larger values of $A$ translate the
curves to the right). Above the value $L=4$ of the  half-width of a stable compacton, wider
compact pulses produce an intersection for each curve (each $A$) with the axis $L=4$, providing a
series of compactons of different heights, like in the numerical experiments described in 
\cite{compacton1,fred,rosenau2}. Below the $L=4$ line, all the curves $L(V)$ approach $L=0$,
towards infinite amplitude, explaining  the instability of the narrower initial data.

Another good example  of the predictive power of the method is exemplified 
in the case of a general convection-nonlinear dispersion
equations, denoted by K(n,m)
\begin{equation}
\eta _t + (\eta ^n)_x +(\eta ^m)_{xxx}=0.
\end{equation}
Compacton solution for any $n\neq m$ are not known in general, except for some particular
cases. In this case we find a general relation among the parameters, for any $n,m$, shown in 
the
fourth and fifth rows. These general relations $L(A,V)$  approach the known relations for
the  exact solutions, in the  particular cases like $n=m$ (fourth row),
$n=m=2$ (third row), $n=m=3$ (first reference in \cite{compacton1}) and $n=3,m=2$;
$n=2,m=3$ (fifth row).  These results can be used to predict the behavior of
solutions for all values of $n,m$. 

In the following, we present  another example of applications of the OSA approach,
related to a new type of behavior of nonlinear systems. Traditional solitons  move
with constant speed on a rectilinear path (except for the  roton \cite{prl} which has a
circular trajectory with constant angular velocity). The speed is usually equal to the
amplitude scaled with a constant. Higher solitons travel faster and there are no 
solitons at rest (zero speed implies zero amplitude). They can travel in  both
directions with opposite signs for the amplitude.
The situation is  different in the case of compactons, which allow also stationary
solutions. When linear and nonlinear disspersion occur simultaneously, like in  the so
called K(2,1,2) equation
$$
u_t +(u^2 )_{x} + (u)_{xxx} +\epsilon (u^2 )_{xxx}=0,
$$
where $\epsilon $ is a control parameter, the OSA yields a dependence of the
form  
$$
L=\sqrt{(\pm A+\epsilon )/(V\pm A)},
$$
which still provides a constant width if $V=\pm A +2\epsilon $.  In this case, the speed is
proportional to the amplitude, but can change its sign even at non-zero amplitude.
Solutions with larger amplitude than a critical one ($A_{crit}=\mp 2\epsilon$)
move to the right, solutions having the  critical amplitude are at rest, and solutions
smaller than the critical amplitude move to the left. This behavior was 
explored in \cite{rosenau1}, too. However, such a switching of the speed is
not necessarily a feature of the nonlinear dispersion. 
A  compacton of amplitude $A$ on the top of a infinite-length KAK solution of 
amplitude $\delta $
\begin{equation}
u(x,t)= A cos ^2 \biggl ( {{x-Vt}\over{4}}  \biggr ) +\delta , 
\end{equation}
is still  a solution of the  K(2,2) equation, with the
velocity given by $V={3 \over 4}\biggl (2 \delta + A  \biggr )$. For $A=-2 \delta$ the
solution becomes an anti-compacton moving together with the  KAK. 
In the case of a slow-scale time-dependent amplitude 
the oscillations in  amplitude can transform into oscillations in the velocity.
The key to such a conversion of oscillations is the coupling between the
traditional nonlinear picture (convection-dispersion-diffusion) and the typical
Schr\"odinger terms.

In Table 2 we present another class of NPDE, namely the dissipative ones.
These equations generalize the linear wave equation (first row) where there is no
typical length of the traveling solutions. The wavelet analysis provides the
correct expression  for the dispersion relation ($V=c$ $\rightarrow$ $k^2=\omega^2/c^2$) with
no constraint on either the amplitude $A$ or on the width $L$.

In the second row we introduce the celebrated Burgers equation which 
represents the simplest model for the convective-dissipative interaction. Dissipative systems
are to a large extent indifferent to how they were initialized, and follow their own intrinsic
dynamics. We provide in the second column an analytic solution of the Burgers equation.
For some special of values of the integration constants ($2C<V^2$, $D=0$) 
the solution becomes a traveling kink
\begin{equation}
u(x,t)=V+\sqrt{V^2-2C} \hbox{tanh}[\sqrt{V^2 -2C} (x-Vt)] .
\end{equation}
By applying the OSA approach to the Burger  equation (third column) we obtain
the same relation between amplitude and half-width, like in the case of the exact solution
eq.(10), providing the velocity is proportional with the amplitude.

In the following we apply the OSA  approach to investigate a 
nonlinear Burgers equation
\begin{equation}
u_t +a (u^m )_x -\mu  (u^k )_{xx}+cu^{\gamma}=0,
\end{equation}
 called quasi-linear parabolic equation \cite{rosenau3}, and used to describe the flow of
fluids in porous media or the transport of thermal energy in plasma. The last term describes
the volumetric absorption (Bremsstrahlung for $\gamma = 1/2$, synchrotron radiation for
$\gamma$ in the range $1.5 - 2$, etc). The second term in eq.(11) describes the convection
process and the coefficient $\beta $ ranges from 0 to 1 in the case of difussion in plasma, and
further to higher values, for unsaturated porous medium in the presence of gravity.
The existence and stability of waves or patterns is strongly dependent on the coefficients $a,
\mu, c, m, k$ and $\gamma $, and at this point the OSA can be
useful again since there is no general analytic solution for eq.(11). The result of the
OSA approach is presented in the third row of Table 2. The typical scale of
patterns depends on the parameters in the equations and the amplitude of the excitations, in a
complicated way. 

However, in order to test OSA again, we found a simple class of  exact solutions when
$c=0$, presented in the  fourt row in Table 2, and expressed as the inverse of a degenerated
hypergeometric function.  In this expression we have ${\cal A}= \mu k V^{\alpha -1 }
((k-1)a^{\alpha })^{-1}$,
$z=(a/V) u^{m-1}$ and $\alpha =(k-1)/(m-1)$.
The asymptotic behavior of the left hand side of the solution given in fourth row, second column,
is described by
$$
\Gamma (\alpha +1) \biggl [ 
(-1)^{\alpha} +{1 \over {\Gamma (\alpha )}}z^{\alpha -1}e^{\alpha }
\biggr ]+{\cal O}(1/z).
$$
If $z$ approaches $+ \infty$ the solution increases indefinitely like an exponential.
For $\alpha >1$ (strong difussion effects), for even $k$ and for even $m$, the traveling wave
$u(x-Vt)$ has a negative singularity towards $-\infty$ at $x+x_0 =\Gamma (\alpha +1)
(-1)^{\alpha}<0$. For $k$ odd there is also a singularity at $x+x_0 >0$. These solutions are
not likely to provide viable physical results. If $k$ is even and $m$ is odd (the singularity
is pushed towards imaginary $x$), or if
$0 < \alpha <1$, the singularity is eliminated and the  solution becomes semi-bounded, like in
the particular situations investigated in the article \cite{rosenau3}. In this case,
OSA provides again the correct relations, since we 
obtaine the special behavior of the solution if the velocity is proportional to the power
$m-1$ of the amplitude $A$. Also, we predict  the space scale of these
semi-compact pulses, namely the length  $L={{\mu k^2 A^{k-1}}\over{V\pm  amA^{m-1}}}$ .

The OSA analysis can be applied in the  case of sine-Gordon equation, fifth row of Table 2.
The solutions with the velocity  proportional with $L^2$ are characterized
through the OSA approach by a transcendental equation in $A$, identical
with the equation fulfilled by the amplitude $A$ of the exact sine-Gordon soliton.

In the sixth row, we present the cubic nonlinear Schr\"odinger equation (NLS3) which has 
a soliton solution. This type of equation is applied in nonlinear optics, elementary
particle physics or in the polaron model in solid state physics \cite{optics,molecular,nuclear}.
Recently, different effects of cluster physics could be explained by using the NLS3 equation.
In the sixth row of Table 2 we present the NLS3 equation together with its one soliton
solution of amplitude $\eta_0$, obtained by the inverse scattering method. In the last column we
also show the relation between the parameters of a localized solution, obtained by OSA. The
equation for $L(A,V)$ is more general than that one fulfilled by the soliton, and hence is
related to more general localized solutions. By choosing the velocity  proportional to the
amplitude, we reobtain the $L\sim 1/A =1/\eta_0 \sim 1/V$ typical relations for the soliton given
in the second column. If a more general solution of the NLS3 equation describes, for example, the
dynamics of some cluster states, or the dynamics of hard spheres in a hard core potential model
$$
-{{\hbar ^2}\over {2m}} \Psi _{xx} +(E-V)\Psi + a \Psi ^3 =0,
$$
then the $L$ parameter gives an estimation for the wavelength
of the wavefunctions, or for the correlation length in a Bose model
$$
L={{\hbar}\over {\sqrt{2m(E-V)+a A^2}}}\simeq {{\hbar}\over {\sqrt{2m(E-V)}}} \ \ 
\hbox{for
small } \ A.
$$
For the general case of a NLS equation of order
$n$ (seventh row), where a general analytical solution is unknown, the method predicts a
special $L=L(A,V)$ dependence, shown in the third column and in Fig. 4. Contrary to third order
NLS, where the dependence of $L$ with $A$ is monotonous for $V=\sim \pm A$ ($n=3$ in Fig. 4), at
higher orders than 3, the $L(A)$ function has  discontinuities in the first derivative. This
wiggle of the function (Fig. 4, $n=4$) holds at a  critical width, possibly producing
bifurcations in the solutions and scales. As a consequence,  initial data close to this
width can split into  doublet (or even triplet, for higher order NLS) solutions, with
different amplitudes. Such phenomena have been put into evidence in several numerical
experiments for quintic nonlinear equations \cite{rosenau2,rosenau3,kart}.

The final example of Table 2 is provided by the Gross-Pitaevski (GP) mean field equation, which
is used to describe the dilute Bose condensate \cite{bose1}. The scalar field (or order
parameter) governed by this equation was shown to behave  in a particle manner, too, since it
can contain topological deffects, namely dark solitons. The space scale $L$ of such solutions
is  important, for both the theory and experiment, since is related to the trap dimensions
and to the scattering length. In the last row of the Table 2 we give one particular solution
of a simplified one-dimensional version of the  GP equation \cite{bose2}
\begin{equation}
i\hbar {{\partial \Psi ({\vec x},t)} \over {\partial t}}=\biggl (
-{{\hbar ^2 \triangle} \over {2m}} +V_{ext}({\vec r}) +{{4\pi \hbar ^2 a} \over {m}}
|\Psi ({\vec x}, t)|^2
\biggr ) \Psi ({\vec x}, t),
\end{equation}
where $a$ is the  s-wave scattering length and $V_{ext}$ is the confining potential. In the
solution provided in the table, the half-width of the exact nonstationary solution
is $L=1/\sqrt{v_c ^2 - p^2}$, where $v_c \def \sqrt {1-aV}$ is the Landau critical velocity, and
$p={\dot q(t)}$ is the momentum associated with the motion of this disturbance.
It is easy to check that the OSA  provides a good match
with this exact solution, and also $L$ fits the correlation length $l_0 = \sqrt{m/4\pi \hbar ^2
a}$. We stress that such estimation of the length is also important in nuclear physics
where one can explain the fragmentation process as a bosonization in $\alpha $-particles, inside
the nucleus. Such systems are coherent if the wavelength associated with the cluster (the
resulting $L$ in the GP equation) is comparable with the distance between the $\alpha
$-clusters.

The more complex the NPDE is, the richer the conclusions of the OSA approach. A good example of
such an analysis is related to the convective-dissipative-dispersive NPDE, example  provided by
the model equation
\begin{equation}
u_t + a(u^m ) _{x} + b (u^k )_{xx}+ c(u^n)_{xxx}=0
\end{equation}
Here $m, k$ and $n$ are integers and the corresponding terms are responsible
of the nonlinear interaction (convective term), dissipation and dispersion \cite{rosenau3}.
The above equation is related to weakly nonlinear phenomena, and it occurs in 
modeling porous medium, magma, interfacial phenomena in fluids (and hence applications to drop
physics), etc. General solutions are difficult to obtain for such a complex equation. 
We show in the following that most of the essential conclusions for the behavior of its solutions
can be obtain, in a simple way, by the OSA. This approach maps this equation
into
\begin{equation}
(amA^{m-1}-V)L^2 -\mu k^2 A^{k-1}L+n^3 A^{n-1}=0,
\end{equation}
Table 3, first row.
The most symmetric case is obtained when either $V=0$ (stationary patterns)
or $V\sim A^{m-1}$. In this situation the condition to have a monotonous dependence
of $L$ as a function of $A$ is $2k=m+n$ which yields a scale structure
\begin{equation}
L=A^{k-m}
\biggl (
{{\mu k^2 \pm \sqrt{
\mu ^2 k^4 -4mn^3 (a-V_0 )
}}\over {2m(a-V_0 )}}
\biggr )\sim A^{k-m},
\end{equation}
where we put $V=mV_0 A^{m-1}$.
The condition $2k=m+n$ is just the condition obtained in \cite{rosenau3} from a scaling 
approach. This condition assures the universality of the corresponding patterns, and it is the
unique case in which $L$ depends on a power of $A$. In any other situation, this dependence is
more complicated and introduces singularities which broke the self-similarity.
In the above cited paper, the author finds out the condition for mass invariance at scaling
transformations as $m=n+2=k+1$. In our case we just have to request the product $AL$ (which
gives a measure of the mass, or volume of the pattern, like in the case of
one-dimensional solitons) to be a constant. This gives the condition $k-m+1=0$ which, together
with the general invariance condition $2k=m+n$, reproduces $m=n+2=k+1$. In  this case we have
patterns characterized by a width
$$
L={{\mu k^2 \pm \sqrt{\mu ^2 k^4 -4mn^3 (a-V_0 )}}\over {2mA(a-V_0 )}}\rightarrow n^3 /A\mu k^2.
$$
If $a\sim V_0$ the width approaches $ n^3 /A\mu k^2$.
In order to make $L$ independen of $A$, like in the compacton case, we need $m=k$, which
together with the first invariance condition $2k=m+n$, yields $m=n=k$. This is the
exceptional case when the dissipative and dispersive processes have the same scaling, resulting
form the invariance of the eq.(13) under the group of scales.
Finaly, if we choose $L\sim V$ we obtain the condition $k+1=2m$ which (together with $2k=m+n$)
is the condition for spiral symmetry and occurence of similarity structures \cite{rosenau3}.

The above comments are not intended to be a complete study of such a complex equation, but the
just a proof of  how many conclusions one can obtain from the simple equation in $A,L$ and $V$,
eq.(14).

The next example is provided by one of the most generalized KdV equation, which is
generated from the Lagrangian \cite{fred}
\begin{equation}
{\cal L}(n,l,m,p)=\int \biggl [
{{\phi_x \phi _t}\over {2}}+\alpha {{(\phi_x)^{p+2}}\over {(p+1)(p+2)}}
-\beta (\phi_x)^m (\phi_{xx})^2 +{{\gamma }\over 2} (\phi _x)^n (\phi _{xx})^l (\phi _{xxx}) ^2 
\biggr ],
\end{equation}
where $\alpha, \beta$ and $\gamma$ are parameters adjusting the relative strentgh of the
interactions, and $n,l,m,p$ are integers. For example, for $\gamma =0$ one re-obtains the K(2,2)
equation, and for $\gamma=m=0, p=1$ one obtains the KdV equation.
The associated Euler-Lagrange equation in the function $\phi_x =u(x,t)\rightarrow u(x-Vt)=u(y)$,
reads after one integration
$$
Vu={{\alpha }\over {p+1}}u^{p+1}-\beta mu^{m-1}(u_y )^2 +2\beta (u^{m} u_y )_{y}
+{{\gamma n}\over {2}} u^{n-1} (u_y )^l (u_{yy})^ 2 
$$
\begin{equation}
-{{\gamma l} \over 2} (u^{n} (u_y )^{l-1}
(u_{yy})^2 )_{y}+\gamma (u^n (u_y )^l u_{yy})_{yy} +C,
\end{equation}
where $C$ is the integration constant. By using the OSA  we obtain
the following important result, expressed in the second row of Table 3: The unique case when such
an equation allows compact supported traveling solutions is when $m=p=n+r$, $C=0$ and $V=V_0
A^{m}$. This result is in full agrement with the variational calculation in \cite{fred}.

Both eqs.(13) and (17) are rather more qualitative than capable of modeling measurable phenomena.
That is why we introduce now a more general model equation, in the form
\begin{equation}
u_t +f(u)_x +g(u)_{xx}+h(u)_{xxx}=0,
\end{equation}
where $f,g$ and $h$ are differentiable functions of the the function $u(x,t)$ itself.
The OSA approach gives the equation
\begin{equation}
-V+f'(A)+{{Ag''(A)+g'(A)}\over L}+{{A^2 h'''(A)+3Ah''(A)+h'(A)}\over {L^2}}=0.
\end{equation}
A general analysis of eq.(19) is difficult, and the best  ways are numerical investigations 
obtained for particular choices of the three functions. We confine ourselves  here only to show
that the class of solutions which have similarity properties are those for which
$V=V_0 f'(A)$. In this case eq.(19) can be reduced to
\begin{equation}
L^2 f' (1-V_0 )+L(Ag'' +g' )+A^2 h''' +3A h''+h'=0,
\end{equation}
case which is presented in the third row of Table 3.
This last relation can be used for different purposes. For example, given a certain
type of dispersion and difusion ($g,h$ fixed), we can estimate for what types of nonlinearity
($f$) the width $L$ will have a given dependence with $A$. Or, if we know for instance
$f(u)=f_0 u^{q_1}$ and $h(u)=h_0 u^{q_2}$, we can ask  what type of diffusion $g$
we need, to have constant scale (width) of the patterns (waves), no matter of the magnitude of
the amplitude $A$. In other words, which is the compatible diffusion term, for given
nonlinearity-dispersion terms, which provides fixed scale solutions.
The result is obtained by integration eq.(20) with respect to $g(u)$ 
\begin{equation}
g(u)=-{{h_0 }\over L} \biggl ( 1+q_2 +{1 \over {q_2 -1}}\biggr ) u^{q_2} -{{Lf_0 (1-V_0
)}\over{q_1 -1}}u^{q_1} +C_3 \hbox{Log} u +C_4, 
\end{equation}
where $C_{3,4}$ are constants of integration. In a similar way one can check the existence of
different other configurations by solving  eq.(20), or more general,  eq.(19).

A last application of this method, occurs if the KdV equation has an
additional term depending on the square of the curvature
\begin{equation}
u_t +uu_x +u_{xxx} +\epsilon (u_{xx}^{2})_{x}=0.
\end{equation}
This is the case for extremely sharp surfaces (surface waves in solids or
granular materials) when the hydrodynamic surface pressure cannot be linearized 
in curvature. Such a new term yields a new type of localized solution fulfilling the relations
$$
L=\sqrt{
{{4\epsilon A}\over {\pm \sqrt{1-8\epsilon  A(A\pm V)}-1}}}.
$$
If we look for a constant half-width solution (compacton of $1/L =\alpha $) we need
a dependence of velocity of the form $V=(1+\alpha ^2 \epsilon /8)A+1/8\epsilon A +\alpha /4$.
There are many new effects in this situation. The non-monoton dependence of the
speed on $A$ introduces again bifurcations of a unique pulse in dublets and triplets.
Also, there is  an upper bound for the amplitude at some critical values of the width. Pulses
narrower than this critical width drop to zero. Such bumps can exist in
pairs of identical amplitude at different widths. They may be related with
the recent observed "oscillations" in granular materials \cite{rosenau1,oscillon}.

The examples presented in Tables 1-3 prove  that the above method provides a reliable
criterion for finding compact suported solutions. The reason this simple
prescription works in so many cases follows from the advantages of wavelet
analysis on localized solutions. We stress that this method has little to do
with the traditional similarity (dimensional) analysis
\cite{compacton1,bona,fred,rosenau1,rosenau2,simil}. In the latter case one obtains
relations among powers of $A, L$ and $V$, not relations with numeric coefficients like those
found in our method.

\section{The frame of KAK pairs}

In the following we investigate the posibility of construction of a nonlinear frame
(an over determined or incomplete basis) by using 
some compact solutions of the K(2,2) equation.
The high stability against scattering of the K(2,2) compactons,   or 
compacton generation from compact initial data, suggest  they may play the role of a nonlinear
local basis. We know from many numerical experiments \cite{bona,fred,electronic},
that  any positive compact initial data  decomposes into a finite series of compactons and
anticompactons. This suggests that an intrinsic ingredient for  a nonlinear basis could be the
multiresolution structure of the solutions, similar with the structure of scaling functions in
wavelet theory.

The compactons given in eqs.(5,6) have constant half-width  and hence
describe a unique scale, which can cover all the space by integer translations. From the point
of view of multi-resolution analysis, the K(2,2) equations act like a
$L$-band filter, allowing only a particular scale to emerge for any given set
of initial condition. To each scale, from zero to infinity, we can associate a K(2,2)
equation with different coefficients. However, the compacton solution is not the
unique one with this property.  For a given K(2,2) equation, we can thus extend the scale from
$L$  to any larger scale. These more general compact supported solutions are still
$C_{2}({\bf R})$ and are combinations of piece-wise constant and piece-wise $\cos ^2$
functions. The simplest shape is given by a half-compacton prolonged with a constant level,
that is a kink solution. The
basis solution is a kink-antikink (KAK) compact supported combination, Fig. 1. Such
kink-antikink pairs of different length, can be associated with other compactons, or KAK pairs,
one on the top of the other
\begin{equation}
{\eta}_{comp+KAK} (x-Vt; \lambda)=
\left\{ \begin{array}{ll}
0 ...\\
{{4V}\over 3} \  cos^{2}\biggl [ {{x-Vt}\over 4} \biggr ],
\ \   -2\pi \leq x-Vt \leq 0 \\
{{4V}\over 3} ,  \ \   0 \leq x-Vt \leq \delta  \\
{{4V}\over 3}+{{4}\over 3}(V'-2V) \  cos^{2}\biggl [ {{x-V't}\over 4} \biggr ],
\ \   \delta \leq x-Vt \leq \delta +4\pi \\
{{4V}\over 3} ,  \ \   \delta+4\pi  \leq x-Vt \leq \lambda \\ 
{{4V}\over 3} \  cos^{2}\biggl [ {{x-Vt-\lambda }\over 4} \biggr ],
\    \lambda \leq x -Vt \leq \lambda +2\pi 
\\ 
0 ... \\
\end{array}
\right.
\end{equation}
where $\delta < \lambda$ characterizes the initial position (at $t=0$) of the top compacton,
with respect to the flat part of the KAK solution. The amplitude $4(V'-2V)/3$ of the compacton,
and the amplitude $4V/3$ of the KAK,  are related to their velocities $V'$ and $V$,
respectively.  The length of the flat part, $\lambda $, is arbitrary. A compound solution is
not stable in time since the different elements travel with different velocities. The total
height of the compacton is $4(V'-V)/3$. Since the higher the amplitude is, the faster the
structure travels, the top compacton moves faster than the KAK, and at a certain moment it
passes the KAK. Because the area  of the solution is conserving, such a compound structure
decomposes into compactons and KAK pairs. Similar and even more complicated constructions can be
imagined, with indefinite number of compactons and KAK's,  if one just
fulfills the $C_3$ continuity condition for the square of the total structure.
Such structures, defined at the initial moment can interpolate any function, playing a similar
role with wavelets or spline bases. 
It has been also proved that the KAK solutions are stable, by using both
a linear stability analysis and  Lyapunov stability criteria.

For a given K(2,2) equation, the compacton solution, eq.(5) and in addition the family of KAK
solutions, eq.(6) can be organized as a scaling functions system. They  act
like a low-pass filter in terms of space-time scales and 
give the opportunity to construct frames of functions from the wavelet model
\cite{wavelet1,wavelet2,wavelet3}. 

With the notation from Appendix 2, and from eq.(32), we have the elements of the frame
$$
\eta _{k,j} (x)=\eta_{kak} (\pi (x-2^{j}Vt-k), 2^{j}-1)|_{t=0},
$$
where $t=0$ means that we neglect the time evolution, but the amplitude is still amplified  with
a factor of $2^j$, in virtue of relation $\eta_{max}=4V/3$. 
We can now expand any
initial data for the K(2,2) equation in this  frame
\begin{equation}
u_{0} (x) = \sum_{k}\sum _{j} C_{k,j} \eta _{k,j} (x),
\end{equation}
and taking into account that
\begin{equation}
\eta _{k,j} \eta _{k',j'}
\left\{ \begin{array}{ll}
\neq 0 & k'=k\cdot 2^{j'-j}, ..., (k+1)\cdot 2^{j'-j}-1 \\
=0 & \mbox{otherwise,}
\end{array}
\right.
\end{equation}
we can show
that the square of the initial data can be linearized by
$$
u^2 (x)=\sum _{k,j} \sum _{j' \geq j} \sum _{k' \in I}C_{k,j}C_{k',j'}
$$
\begin{equation}
\times \biggl (
\sum_{i_1 =0}^{1} \sum_{i_2  =0}^{1}... \sum_{i_{j'-j}=0}^{1} \eta _{\sigma (i_1
,i_2 , ..., i_{j'-j}),j'}
\biggr ) \eta _{k',j'},
\end{equation}
where $I$ is the range of $k'$ described in the first line of eq. (25), and 
$$
\sigma (i_1 , i_2 , ..., i_{j'-j}) =
\sum_{l=1}^{j'-j} i_l 2^{j'-j'l+{{(j'-j)(j'-j+1)-l(l+1)}\over 2}}
$$
\begin{equation}
+k2^{(j'-j)j+{{(j'-j)(j'-j+1)}\over 2}}.
\end{equation}
From this relation and from eq.(25) we notice that in eq.(27) the unique nonzero terms are
those for which $\sigma (i_1 , i_2 ,..., i_{j'-j} )=k'$ with $k' \in I$. Hence the initial data
is expanded in different scales at different translations. The translations are mutually
orthogonal so they do not give a contribution to the square. In the  multiplication of  two
different scales in the expression of the square, we reduce the wider scale in
terms of linear combination of the narrower ones, by using the two-scale equation,
eq.(31) in Appendix 2. All the nonzero  terms in this product
are of the order  $(2^{-j}-1)/(2^{-j'}-1)\simeq 2^{j'-j}$. This number, given
by the number of solutions of the equations $\sigma (i_1 , i_2 ,..., i_{j'-j} )=k'$, 
with $k' \in I$, is much smaller than the initial number of terms, where from the advantage of
the frame.  This is the advantage of treating nonlinear problems with a basis that
has a scale criterion.

Another application of the KAK basis occurs when one needs to understand the dynamics of
the initial data for the K(2,2) equation. Many numerical simulations show that
in the case when the width of the initial data is larger that $L_{compacton}$,
the initial shape decomposes into a finite number of compactons having the same width and
different amplitudes \cite{compacton1,fred,rosenau2,rosenau3}.
If we take such an initial pulse, Fig. 5a, and we want to see its time evolution untill the
break up process, we have first to expand it into the KAK frame, Fig. 5a.
Then we just let the system develope and the different KAK's move with their correspon
velocity, eq.(23), providing the new shapes, Fig. 5b and 5c, at different moments. We notice a
good agreement between these theoretical calculation and the above quoted numerical experiments.
At this stage we can not yet predict the way that the  KAK breaks back into compactons. However,
we notice that the area of such structures in the phase space is the Poincare invariant and
should give a hint towards the breaking process. This invariant is nothing that the $u^3$
(denoted
$D_2$ invariant in [4]) invariant of the K(2,2) equation.

\section{Comments and conclusions}

First, we make a general statement concerning the compact solutions of one-dimensional NPDE.
A one-dimensional dynamical model is described by a general NPDE  equation 
$\partial _t u ={\cal O}(x,{\partial }_{x} ) u$
where ${\cal O}$ is a nonlinear differential operator. By taking into account {\it
only} traveling solutions, this NPDE reduces to a nonlinear ordinary differential equation in
the  coordinate $y =x-Vt$ for an arbitrary velocity
$V$. If $u(y )$ is a compact  supported solution it follows that it
is not unique with respect to fixed initial compact data. Indeed, if we choose the initial
data such that the function  and its derivatives (up to the
requested order) are zero in the neighborhood of a  certain point $y _0$ of the $y $ axis,
these conditions can be fulfilled by any  translated version of
a compact supported particular solution, placed  everywhere
on the axis outside this neighborhood. Consequently, for such initial data, the
solution is not unique. This result shows that the compact supported property of the initial
data and of the solution implies its non-uniqueness.

Since we can transform the NODE into a nonlinear differential
system of order one
\begin{equation}
{\vec U}_y ={\vec F}(y ,
 {\vec U}), \ \ \ {\vec U}=(u, u_y , ...) , 
\end{equation}
we can apply the fundamental theorem of existence and uniqueness to solutions
of eq.(28), for  given initial data ${\vec U}(y _0 )={\vec U}_0$.
If the function ${\vec F}$ in eq.(28) fulfills the Lipschitz condition  (its relative
variation is bounded) than, for any initial condition, the solution is unique \cite{hure}.
Since  any linear function is analytic and hence Lipschitz, we  conclude
that only nonlinear functions ${\vec F}$ allow the existence of
compact supported solutions. Thus, a compact soliton 
implies non-uniqueness in the underlying NPDE, which implies non-Lipschitzian structure
of the NPDE and hence the existence of nonlinear terms.

In this paper we  introduce new physical applications for
wavelets, that is the study of localized solutions of  nonlinear partial differential
equations. The existence of compactons underlines a common feature of NPDE, discrete
wavelets,  and also finite differences equations.  We propose a new scale approach for
the similarity analysis and clasiffication of soliton solutions, without the need of solving
the corresponding NPDE. Also, we proved that starting from any unique soliton solution of a
NPDE, we can construct a frame of solutions organized  under a multiresolution criterium.
This approach provides the possibility of constructing a
nonlinear basis  for NPDE. We show that frames
of self-similar functions are related with 
solitons with compact support.   In addition, we notice the evidence
that compactons fulfil both characteristics of solitons
and wavelets, suggesting possible new applications.
Such a unifying direction between nonlinearity and self-similarity, can bring
new applications of wavelets in cluster formation, at any scale, from
supernovae through fluid dynamics to atomic and nuclear systems. The scale approach
can be applied with success to the physics of droplets,
bubbles,  patterns, fragmentation, fission and
 fusion.

\vskip 1cm 
Supported by the U.S. National Science Foundation through a regular grant,
No. 9970769, and a Cooperative Agreement, No. EPS-9720652, that includes
matching from the Louisiana Board of Regents Support Fund.

\section{Appendices}

\subsection{Appendix 1}

We base our proof on the similar proposition in \cite{ijmpe}, excepting that here
we use Gaussian filtering instead of Morlet.
We start form the  the discrete   wavelet expansion of the signal $u(s)$ 
given in eq.(2)
$$
u(s)=\sum_{j}\sum_{k}C_{j,k}\Psi  (2^j s -k)=\sum_{j,k}C_{j,k}\Psi _{j,k}(s),
$$
in terms of integer translations ($k$) and dyadic dilations ($2^j$) of the $\Psi$
wavelet. In order to reduce the number of scales needed, the range of the summation
should be choosen as a function of $s$.
We use in the following the asymptotic formula describing the pointwise
behavior of the Gaussian wavelet series around a point $s_0$ of interest \cite{approx}.
For a chosen $s_0$ and scale $j$, there is only one $k$ and $|\epsilon | \leq 1$
such that the support of the corresponding $\Psi _{j,k}$ contains this point,
$k=2^j s_0 +\epsilon$. We can express the solution and its derivatives in a neighborhood 
of this point 
$$
u(s_0) \approx \Psi (-\epsilon )\sum_{j}C_{j,2^j s_0 +\epsilon}\equiv 
\sum_j u_j (s_0 ) ,
$$
\begin{equation}
u_{x}(s_0) \approx -i \Psi (-\epsilon )\sum_{j}2^j  C_{j,2^j
s_0 +\epsilon}=-i \sum_j 2^j u_j (s_0 )  ,
\end{equation}
for $\epsilon  $ being chosen enough large compared to one, and where $u_j (s_0 )=\Psi
(-\epsilon ) C_{j,2^j s_0 + 
\epsilon}\approx \Psi (0) C_{j,2^j s_0 }$... Since the coefficient 
$1/ 2^j
$ represents the scale for each $\Psi _{j, k}$ Gaussian wavelet, we can define it as
a characteristic half-width
$L_{j}$. Also, we finally have for the $n$-th order derivative in $s_0$
\begin{equation}
u_{xx... \ x}(s_0 )\approx \sum_{j} {{u_{j}(s_0 )}\over {L_{j}^{n}}}.
\end{equation}
Eq.(30) is the multi-scale generalization
of the simpler formula $ii$ in eq(1). With eq.(30) in hand we can investigate the structure of
hypothetic soliton solutions of NPDE, by choosing $s_0$ in the neighborhood of the maximum
value of the solution,  $u(s_0)=A$.  Around this maximum, such solutions can be described
very well by a unique scale $L$, and hence the solution and its derivatives can be
approximated  with the corresponding dominant term, by  the substitutions in eq.(1).

\subsection{Appendix 2}

For the sake of simplicity we will renormalize the coefficients of the
K(2,2) equation such that the support of the simple compacton is one.
That is, we take $\eta _c (x,t)=\eta _{kak}(\pi (x-Vt),0)$ on the interval $|x-Vt|$
in $[-1/2,1/2]$. We construct a multiresolution approximation of $L^{2}({\bf R})$, that is an
increasing sequence of closed subspaces $V_j$, $j\in {\bf Z}$, of $L^{2}({\bf
R})$ with the following properties \cite{wavelet2,wavelet3}
\begin{enumerate}
\item
The $V_j$ subspaces are all disjoint and their union is dense in
$L^{2}({\bf R})$.

\item
For any function $f \in L^{2}({\bf R})$ and for any integer $j$ we have
$f(x) \in V_j $ if and only if $D^{-1}f(x) \in V_{j-1}$ where $D^{-1}$ is an
operator that will be defined later.

\item
For any function $f \in L^{2}({\bf R})$ and for any integer $k$, we have
$f(x) \in V_0 $ is equivalent to $f(x-k) \in V_0$.

\item
There is a function $g(x) \in V_0$ such that the sequence $ g(x-k)$  with
$k\in {\bf Z}$ is a Riesz basis of $V_0$.

\end{enumerate}
In the case of compact solutions of K(2,2) of unit length, we chose for the
space $V_0$ that which is generated by all translation of $\eta _c$ with any
integer $k$. The subspaces $V_j$ for $j \geq 0$ are generated by all integer
translations of the compressed version of this function, namely, by
$\eta_{kak}  (2^{j}\pi(x-Vt),0)$. The subspaces $V_j$ for $j\leq 0$
are generated by all integer translations of the KAK solution of length $\lambda 2^j -1$. For
example, $V_{-1} $ is generated by $\eta_{kak} (\pi (x-2^{j}Vt),0)$. The spaces $V_j, j\geq
0$ are all solutions of K(2,2); the others are not.  The function $g(x)$ is taken to be 
$\eta_{kak}
(\pi (x-Vt),0)$. It is not difficult to prove that these definitions fulfill restrictions
one, three, and four. As for the second criterion, we define the action of the operator 
$D^{-1} f(x) = f(2x)$ if $f(x) \in V_{j}$ with a $j$ positive integer, and
$D^{-1}\eta_{kak} (\pi2^{j}(x-2^{j}Vt),2^{-j}-1)=\eta_{kak} (\pi 2^{j}(x-2^{-j+1}Vt), 2^{-j+1}-1
)$ for negative
$j$. In conclusion, we construct a frame of functions made of contractions of compactons
 and sequences of KAK solutions. We can write the corresponding two-scale
equation which connects the subspaces 
\begin{equation}
\eta_{kak} (\pi (x-Vt), 1)=\eta_{kak} (\pi (x-Vt),0)+\eta_{kak} (\pi (x-Vt-1),0).
\end{equation}
We will denote generically by $\eta _{k,j}$ the elements of this frame, that is
\begin{equation}
\eta _{k,j} (x)=\eta_{kak} (\pi (x-2^{j}Vt-k), 2^{j}-1)|_{t=0},
\end{equation}
where $t=0$ means that we neglect the time evolution, but the amplitude is still amplified  with
a factor of $2^j$, in virtue of relation $\eta_{max}=4V/3$.

\vfill
\eject

\vfill
\eject


\vskip 1cm
\centerline{Figure Captions}
\vskip 1cm
\begin{itemize}

\item
Fig. 1

Examples of compactons, kink-antikink pairs, and mixed solutions of the K(2,2) equation,
together with their velocities.

\item
Fig. 2

A finite series of K(2,2) compactons emerging from  initial compact data of width
larger than of the compacton.

\item
Fig. 3

The half-width $L$ versus velocity $V$ for the K(2,2) equation, for different amplitudes A.
Widths larger than $L_{compacton}=4$ behave different than
narrower widths, with $L<4$. Amplitude increases from left to right, in the range
0.01-0.85.

\item
Fig. 4

The half-width $L$  versus amplitude $A$, for the third ($n=3$), forth ($n=4$) and fifth
($n=5$) order NLS equation, in two  $V=\pm A$ cases. We notice that the higher order ($n>3$) NLS
equations have bifurcations.

\item
Fig. 5

The dynamical evolution of the expansion in the KAK basis of a finite initial pulse of width $L$
much larger than a compacton, drawn for three moments if time.

\end{itemize}

\vfill
\eject

\begin{table}[t]
\caption{Traveling localized solutions for nonlinear dispersive equations.\label{tab:exp}}
\vspace{0.2cm}
\begin{center}
\begin{tabular}{cccc}
\hline \\
NPDE & Analytic solution and the && OSA \\
&relations among parameters & &approach\\
\hline \\

$u_{t}+6uu_{x}+u_{xxx}=0$ & $A~\hbox{sech} ^2 {{x-Vt} \over L};$ \ \ \  $L=\sqrt{2 / A}$, && 
$L=|V\pm 6A|^{-1/2}$\\
 &  $V=2A$ &&If $V\sim A$, $L\sim A^{-1/2}$ \\ 
\hline \\
$u_{t}+u^2u_{x}+u_{xxx}=0$ &   $A~\hbox{sech } {{x-Vt} \over L};$ \ \ \  $L=1/A$, & &
$L=|V \pm 6A^2 |^{-1/2}$ 
\\
&  $A=\sqrt{V}$  &&If $V\sim A^2 ,L\sim A^{-1}$\\
\hline \\
$u_{t}+(u^2)_{x}+(u^2
)_{xxx}=0$  &  $A \hbox{cos} ^2 {{x-Vt}\over L}$, \ \ \hbox{if} \ \ 
$|(x-Vt)/4|\leq \pi /2 $;
 & & $L=\biggl ( {{8A}\over
{|V
\pm 2A|}}\biggr ) ^{1/2}$  \\  
 & L=4 &&   \\
\hline \\
$u_t +(u^n )_x +(u^n )_{xxx}=0$  & $\biggl [ A cos^{2}\biggl ( {{x-Vt} \over {L}}
\biggr ) \biggr ] ^{1 \over {n-1}}$, \ \  if $|x-Vt| \leq {{2n\pi}\over{n-1}}$ & &
$L=\biggl ( {{n(n^2 +1)}\over{\alpha \pm n}}\biggr )^{1/2}$ \\
&and 0 else; & & \\
&$L={{4n}\over {(n-1)}},$ \ \ $A={{2Vn}\over {n+1}} $ && if $V=\alpha A^{n-1}$ \\ 
\hline \\
 $u_t +(u^n )_x +(u^m )_{xxx}=0$ & unknown 
&  & 
$L=\biggl ({{{n(n^2 +1) A^{n-1}} \over {V \pm mA^{m-1}}}}\biggr )^{1/2}$
\\  
$n\neq m$ &in general &
 & \\ 
\hline \\
\end{tabular}
\end{center}
\end{table}

\vfill
\eject

\begin{table}[t]
\caption{Traveling localized solutions for nonlinear diffusive equations.\label{tab:exp}}
\vspace{0.2cm}
\begin{center}
\begin{tabular}{cccc}
\hline \\
NPDE & Analytic solution and the & OSA \\
&relations among parameters & approach\\
\hline \\
$u_{xx}-{1 \over  {c^2}} u_{tt}=0$ & $\sum C_k e^{i(kx\pm \omega t )}$;  & 
$V=c$ \\
 &$k^2 = \omega ^2 /c^2$ & $ A, \ L$   arbitrary  \\
\hline \\
$u_t + uu_x - u_{xx}=0$ & $ \sqrt{C -V^2} \hbox{tan}(\sqrt{C -V^2}{{x-Vt}\over {2}} +D)$
 & $L=(A \pm V)^{-1}$ \\
 & $+V$ & If $V\sim A$,  $L\sim 1/A$\\
\hline \\
$u_t +a (u^m )_x -\mu  (u^k )_{xx}+cu^{\gamma}$ & only particular cases
\ \    & $cA^{\gamma} L^2 + (V\pm amA^{m-1})L$ \\
=0 &known & $\pm \mu k^2 A^{k-1}=0$   \\
\hline \\
$u_t +a (u^m )_x -\mu  (u^k )_{xx}=0$ & $ -{\cal A} z^{\alpha }\ _{1}F_{1}(\alpha , \alpha +1 ,
z) = x+x_0$
\ \    &$L={{\mu k^2}\over{am-\alpha }}A^{k-m}$, \\
 & & if $ V=\alpha A^{m-1}$   \\
\hline \\
$u_{xt}-\sin u=0$ &  $A~\tan^{-1}\gamma ~e^{{x-Vt}\over {L}}$ \ \ \  &
$\pm {{VA}\over {L^2 }}=sin A$ \\ 
& &    If $V=L^2, A=sinA$ \\ 
\hline \\
$i\Psi_{t}+\Psi _{xx}+2|\Psi |^2 \Psi =0$  & $\eta_0 e^{i(\omega t + kx)} sech[ \eta_0 
(x-Vt)]$;
\ \   &  $L={{\pm V \pm \sqrt{|V^2 - 4 A^2 |}}\over {2A^2 }}$ \\
&$ L=1/ \eta_0$&  If $A\sim V, \ \ L=1/A$ & \\ 
\hline \\
$i\Psi_{t}+\Psi _{xx}+|\Psi| ^{n-1}\Psi=0$  & unknown in general    
&$L={{\pm V \pm \sqrt{|V^2 - 4 A^n |}}\over {2A^n }}$ \\
 & &   \\ 
\hline
$i\Psi_t=-{1 \over 2}\triangle \Psi $ &
$ip+\sqrt{v_{c}^{2}-p^2}\times  $  & 
$L=(aA^2 \pm V -1)^{-1/2} $\\
$+[a|\Psi|^2 +V(x) -1]\Psi$  & $\hbox{tanh}[a\sqrt{v_{c}^{2}-p^2}(x-q(t) )]$ 
& If $V\sim \pm 1, L\sim 1/(A\sqrt{a})$ \\ 
\hline
\end{tabular}
\end{center}
\end{table}

\vfill
\eject

\begin{table}[t]
\caption{Traveling localized solutions for  dissipative-dispersive equations.\label{tab:exp}}
\vspace{0.2cm}
\begin{center}
\begin{tabular}{cccc}
\hline \\
The NPDE equation & OSA approach \\
\hline \\

$u_t + a(u^m ) _{x} + b (u^k )_{xx}+ c(u^n)_{xxx}=0$; & 
$L=A^{m-k} \cdot {{\mu k^2 \pm \sqrt{\mu ^2 k^4 -4mn^3
(a-V_0 )}}\over {2m(a-V_0 )}}$ \\
&if  $V=mV_0A^{k-1}$  \\
\hline \\
$Vu={{\alpha }\over {p+1}}u^{p+1}-\beta mu^{m-1}(u_y )^2 +2\beta (u^{m} u_y )_{y}$ &  
$2L^{l+4}((n+l+1)V_0 -\alpha) $\\ 
$+{{\gamma n}\over {2}} u^{n-1} (u_y )^l (u_{yy})^ 2 -{{\gamma l} \over 2} (u^{n} (u_y
)^{l-1}(u_{yy})^2 )_{y}
$&$-2L^{l+2}(l+n+1)(l+n+2)\beta$ \\ 
$+\gamma (u^n (u_y )^l u_{yy})_{yy} +C$& $-(l+n+1)(2+2n^2 +3l+l^2 +n(5+3l))\gamma$ \\ 
&if $C=0$, $V=V_0 A^{m}$ and $m=p=n+l$ \\
\hline \\
$u_t +f(u)_x +g(u)_{xx}+h(u)_{xxx}=0$  &  $L=-\biggl [ g'+Ag''\mp \biggl ( (Ag''+g')^2 $ \\ 
& $ -4f'(1-V_0 )(A^2 h'''+3Ah''+h') \biggr ) ^{1/2} \biggr ] $ \\
& $ \times (2A^2 h'''+6Ah''+2h')^{-1}$ \\ 
& if  $V=V_0 f'(A)$ \\
\hline \\
\end{tabular}
\end{center}
\end{table}

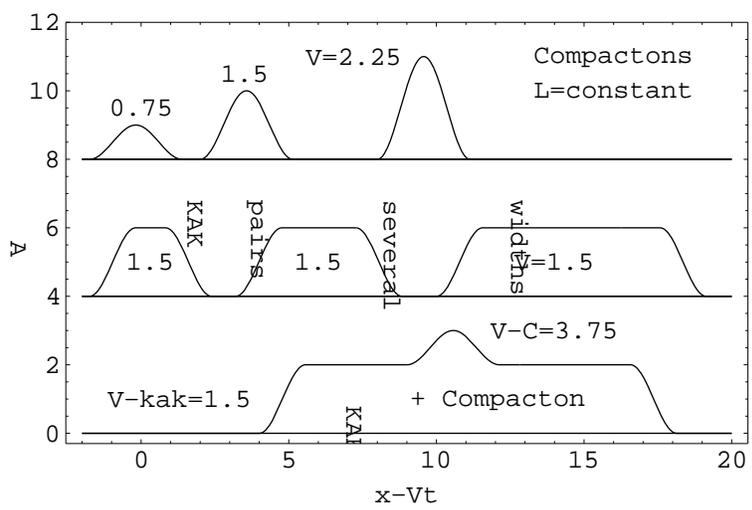
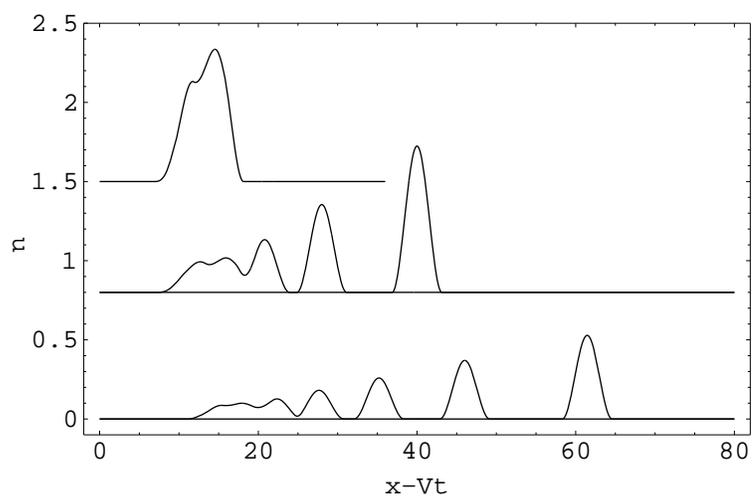
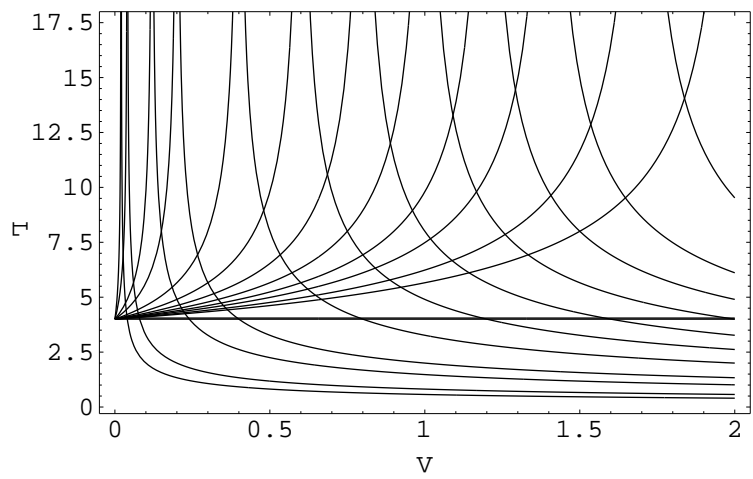
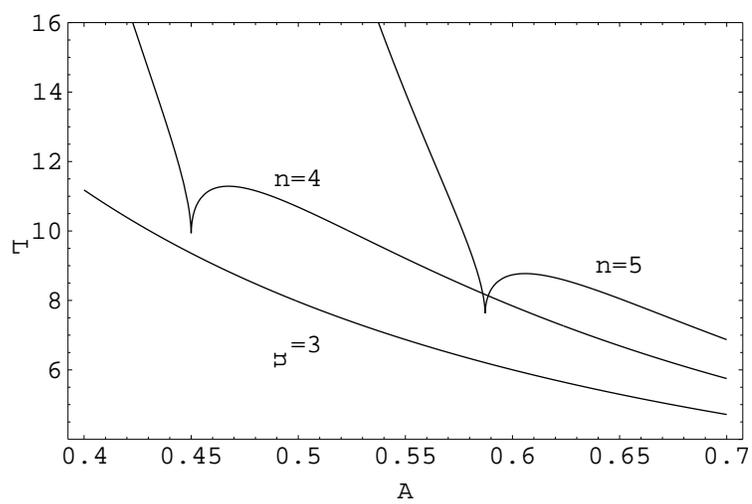
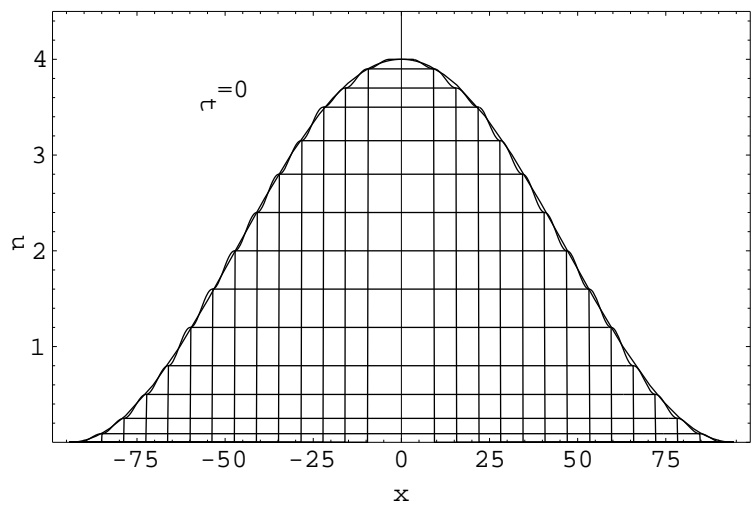
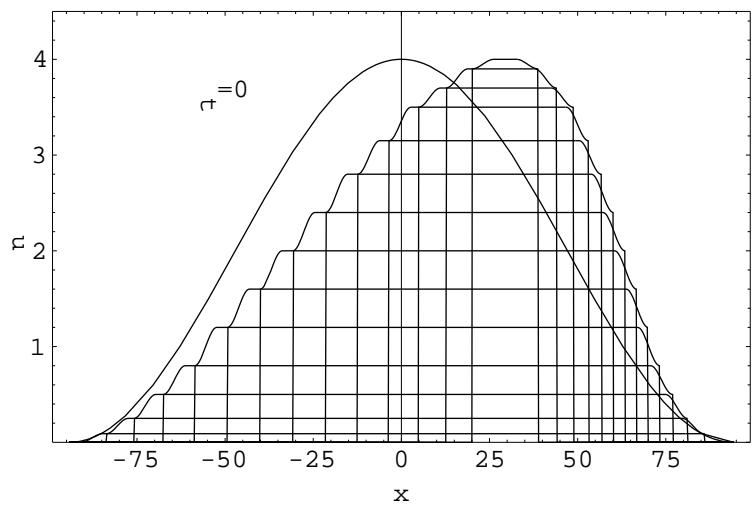
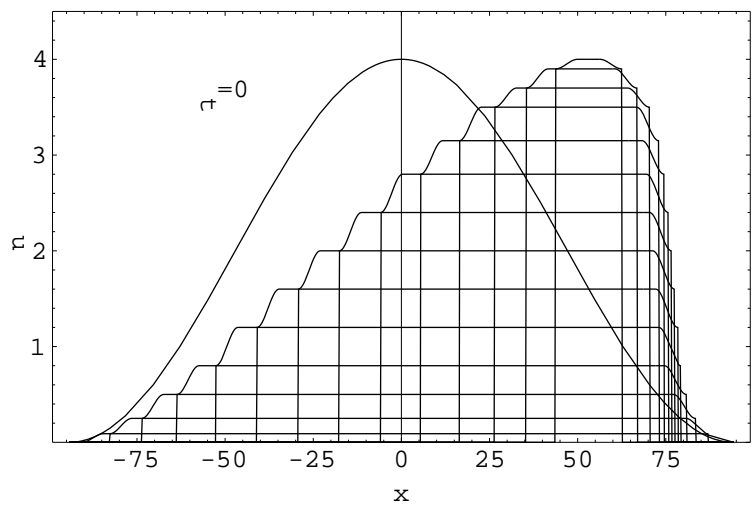
\end{document}